\documentclass[12pt]{iopart}

\usepackage{booktabs}
\usepackage{comment,mfirstuc}
\usepackage{graphicx}
\usepackage{grffile}
\usepackage{algorithm}
\usepackage{algpseudocode}
\usepackage{hyperref}
\usepackage{soul}
\usepackage{listings}
\usepackage{multirow}
\usepackage{dcolumn}
\usepackage{bm}
\usepackage{placeins}
\expandafter\let\csname equation*\endcsname\relax
\expandafter\let\csname endequation*\endcsname\relax
\usepackage{amsmath}
\usepackage{amssymb}
\usepackage{mathtools}
\usepackage[bottom=1.0in]{geometry}
\usepackage{cite}
\usepackage{lineno}
\RequirePackage{xcolor}
\usepackage[misc]{ifsym}
\usepackage{lipsum} 
\usepackage{subcaption}
\usepackage{tcolorbox} 
\usepackage{stfloats} 
\usepackage{siunitx}
\DeclareSIUnit\eVperc{\eV\per\clight}
\DeclareSIUnit\clight{\text{\ensuremath{c}}}

\usepackage{ulem}

\usepackage{xspace}

\usepackage{colortbl}
\usepackage{array}
\newcolumntype{P}[1]{>{\centering\arraybackslash}p{#1}}
\newcolumntype{M}[1]{>{\centering\arraybackslash}m{#1}}

\hypersetup{ 
    pdfnewwindow=true,      
    colorlinks=true,       
    linkcolor=blue,         
    citecolor=blue,        
    filecolor=blue,      
    urlcolor=blue        
}  

\algblock{Input}{EndInput}
\algnotext{EndInput}
\algblock{Output}{EndOutput}
\algnotext{EndOutput}

\def\be{\begin{eqnarray} &&} 
 
\def\ee{\end{eqnarray}}

\makeatletter
\newcommand{\mainmatter}{%
  \setcounter{footnote}{0}%
  \patchcmd{\@makefntext}{\fnsymbol}{\arabic}{}{}%
  \patchcmd{\@thefnmark}{\fnsymbol}{\arabic}{}{}%
  \def\@makefnmark{\textsuperscript{\arabic{footnote}}}
  \long\def\@makefntext##1{\parindent 1em\noindent
        \hb@xt@1.8em{%
            \hss\@textsuperscript{\normalfont\@thefnmark}}##1}%
}
\makeatother

\newcommand{\addComment}[2]{
  \expandafter\newcommand\csname #1\endcsname[1]{{\bf \color{#2} \capitalisewords{#1}:\,##1}}
  \expandafter\newcommand\csname #1cor\endcsname[2]{{\color{#2} \capitalisewords{#1}:\,\st{##1}{\bf ##2}}}
  \expandafter\newcommand\csname #1color\endcsname{#2}
}
\addComment{cris}{blue} 
\addComment{james}{red}
\addComment{cole}{purple}

\begin{document}

\providecommand{\keywords}[1]
{
  \small
  \textbf{Keywords:}  {\color{blue}#1
  }
}

\title[\scriptsize{GPT-Based Fast Simulation of CLAS12 Detector Hits via Conditional Autoregressive Generation}]{GPT-Based Fast Simulation of CLAS12 Detector Hits via Conditional Autoregressive Generation} 

\author{C. Granger$^{1,\star}$, J. Giroux$^{1,\star}$, R. Tyson$^{3}$, M. Ungaro$^{4}$, C. Fanelli$^{1,2,\star}$} 

\address{
$^{1}$ William \& Mary, Department of Data Science, Williamsburg, VA 23185, USA\\
$^{2}$ William \& Mary, Department of Physics, Williamsburg, VA 23185, USA\\
$^{3}$ University of Glasgow, School of Physics and Astronomy, Glasgow G12 8QQ, UK\\
$^{4}$ Thomas Jefferson National Accelerator Facility, Newport News, VA 23606, USA\\
$^{\star}$ Author to whom correspondence should be addressed.
}

\ead{{\color{blue}
cjgranger@wm.edu,
jgiroux@wm.edu,
cfanelli@wm.edu
}}

\vspace{10pt}
\begin{indented}
\item[]\today
\end{indented}

\begin{abstract}
Modern particles physics experiments have demonstrated an increasing need for fast, high-fidelity detector simulation as detector components have improved and subsequent computational requirements approach the limits of available resources. Recently, deep generative models have emerged as a promising alternative to traditional Monte-Carlo methods, with recent works drawing inspiration from large language models (LLMs) and self-supervised next-token prediction methods. In this work, we present an application of a GPT-style autoregressive transformer as a fast surrogate model for the calorimeter inside the CLAS12 experiment at the Thomas Jefferson National Accelerator Facility. The model is conditioned on incident momentum and generates realistic detector hits autoregressively across all nine calorimeter layers as sequences of strip, ADC, and TDC tokens. We demonstrate that the model faithfully reproduces hit multiplicity, spatial distributions, energy deposits, and the energy-momentum response of the electromagnetic calorimeter. The generator achieves inference rates exceeding 700 events per second on a single GPU, providing a substantial speedup over traditional Geant4-based simulations while maintaining physics fidelity essential for high-luminosity experimental programs.
\end{abstract}


%
%
%
%
%

\mainmatter

\section{Introduction}\label{sec:intro}

Accurate detector simulation is a cornerstone of modern nuclear and particle physics experiments, enabling the comparison of theoretical predictions with experimental measurements. At the Thomas Jefferson National Accelerator Facility, the CLAS12 spectrometer~\cite{clas12} 
is located in Hall B of the Continuous Electron Beam Accelerator Facility (CEBAF), designed to study the structure of nucleons and nuclei through inclusive, semi-inclusive and exclusive reactions. A critical component of CLAS12 is its electromagnetic calorimetry system, which consists of three sampling calorimeters, the Pre-shower Calorimeter (PCAL) and the two layers of Electromagnetic Calorimeters (ECIN and ECOUT), used to detect neutral particles and identify electrons via electromagnetic shower reconstruction across nine detector layers.

Traditional \textsc{Geant4}-based~\cite{geant4} full simulations, while highly accurate, are computationally expensive, often requiring orders of magnitude more CPU time than data reconstruction. As experimental programs scale to higher luminosities and larger datasets, fast simulation methods that preserve physics fidelity become essential. 

In recent years, deep generative models have emerged as promising alternatives for fast detector simulation~\cite{fastsim_review, adelmann_2022, Software_CMS, CERN-LHCC-2022-005}. In the past, various deep-learning methods such as Generative Adversarial Networks (GANs) \cite{Paganini_2018,Paganini_2018_accel,Chekalina_2019,Erdmann_2019,carminati2017calorimetry,Belayneh_2020,buhmann2021getting,rehm2021validationdeepconvolutionalgenerative,khattak2021fastsimulationhighgranularity,Bieringer_2022,Rogachev_2023,diefenbacher2023newanglesfastcalorimeter,Giannelli_2024,Simsek_2024}, Variational Auto-encoders (VAEs) \cite{abhishek2022calodvaediscretevariational,cresswell2022calomanfastgenerationcalorimeter,hoque2024caloqvaesimulatinghighenergy,liu2024calovqvectorquantizedtwostagegenerative,smith2024fastmultigeometrycalorimetersimulation, Alex_May_2026, cosso2026calorimetershowersuperresolutionconditional}, normalizing flows \cite{Krause_2023,krause2023caloflowiifasteraccurate,Diefenbacher_2023,Buckley_2024,Pang_2024,Erdmann_2025}, and diffusion models \cite{Mikuni_2022,Buhmann_2023_clouds,amram2023denoisingdiffusionmodelsgeometry,diefenbacher2025refiningfastcalorimetersimulations,Mikuni_2024,buhmann2024calocloudsiiultrafastgeometryindependent,Kobylianskii_2024,Favaro_2025,Buss_2026,raikwar2025generalisablegenerativemodelmultidetector,buss2025caloclouds3ultrafastgeometryindependenthighlygranular,favaro2025fastaccurateprecisedetector,buss2026allshowersmodelcalorimetershowers} have been utilized as efficient surrogate models for simulating calorimeter response. More exhaustive collections of these frameworks can be found in \cite{hepmllivingreview}.
In this work, we present a conditional autoregressive transformer trained to generate realistic CLAS12 calorimeter hit patterns for photon-induced electromagnetic showers. The model is conditioned on the incident particle momentum and generates detector hits autoregressively as a sequence of strip, ADC, and TDC tokens across all calorimeter layers. 
The choice of this transformer-based architecture is motivated by its successful application to both calorimetry \cite{cardona2026generalizable,birk2025omnijet,birk2026spadesplitanddelayembeddingsautoregressive} and Cherenkov-detector reconstruction \cite{giroux2026towards,fanelli2026application}.
For the CLAS12 calorimeter system, this formulation is particularly well suited to the variable-length, multi-layer structure of electromagnetic shower deposits. It can naturally capture correlations among hit multiplicity, strip position, and deposited energy, which are essential features of realistic photon showers in the CLAS12 detector.

\section{Dataset and Detector Description}

The CLAS12 electromagnetic calorimeter system consists of three sampling calorimeters arranged in successive layers: the Pre-shower Calorimeter (PCAL) and the two layers of the Electromagnetic Calorimeter (ECIN and ECOUT)~\cite{clas12_ec}. Each calorimeter is constructed from alternating layers of scintillator strips and lead absorber, with strips oriented in three stereo views --- U, V, and W --- rotated $120^\circ$ relative to one another, yielding nine readout layers in total. This geometry enables three-dimensional shower reconstruction by correlating strip hits across the three views within each calorimeter. When a photon enters the calorimeter, it initiates an electromagnetic shower whose energy and position are reconstructed from the pattern of strip addresses (strips), pulse heights (ADC), and timing information (TDC) recorded across these layers.

The training dataset consists of 1,144,000 simulated photon events produced with the dedicated \textsc{Geant4} framework for CLAS12, \textsc{GEMC}~\cite{gemc}. Each photon is represented by hit-level calorimeter information across all nine strip layers. For each event, up to 20 hits per layer are recorded, giving arrays of shape $(N_\text{events},\, 9,\, 20)$ for strip address, ADC value, and TDC value respectively. The three-momentum of the incident photon $(p_x, p_y, p_z)$ is stored as the conditioning input for each event. Empty hit slots are zero-padded, reflecting the variable multiplicity of shower deposits across layers.

\section{Architecture}\label{sec:architecture}

\begin{figure}[H]
    \centering
    \includegraphics[trim={5mm 25mm 15mm 0mm},clip,width=1.0\linewidth]{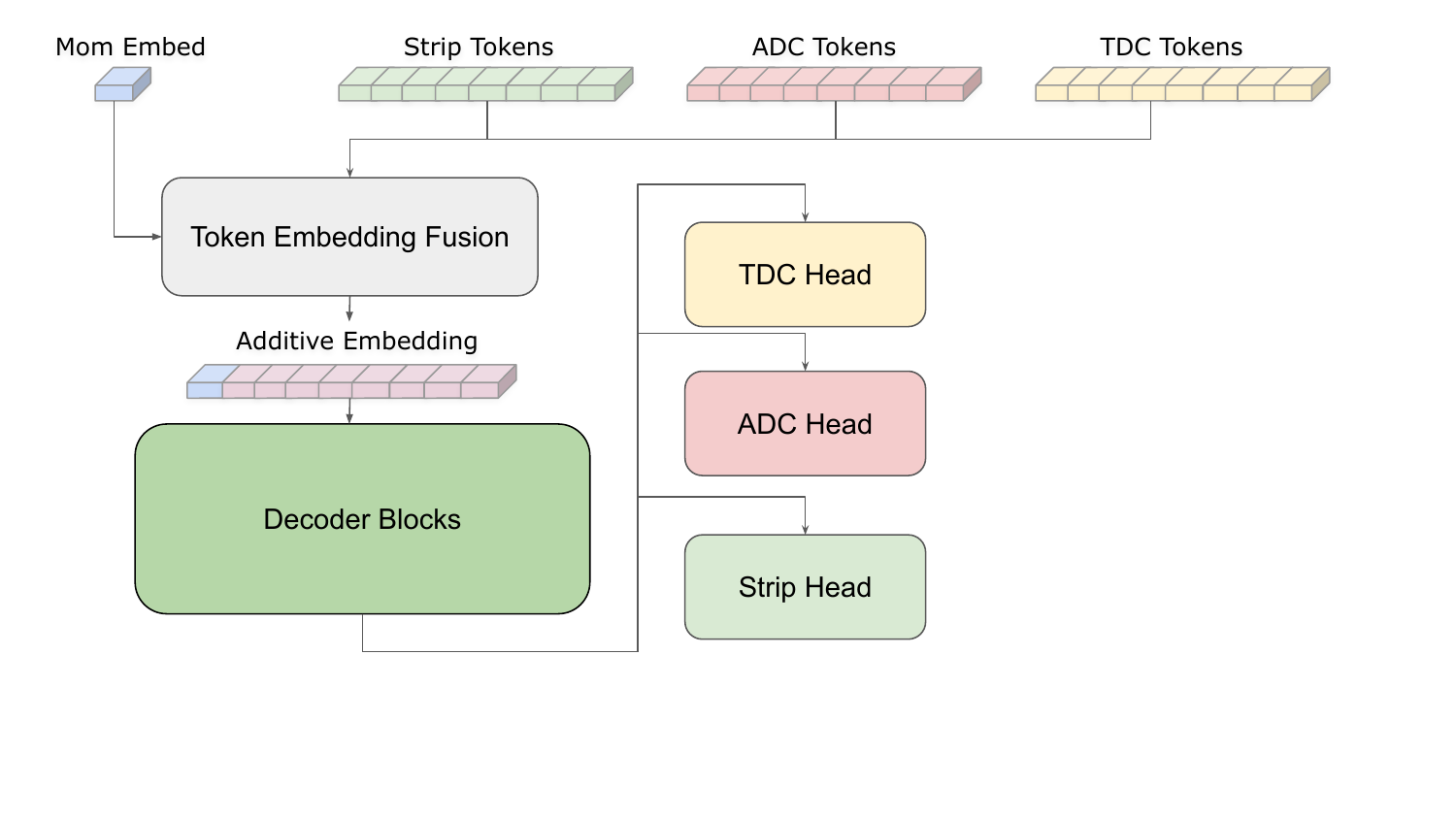}
 \caption{\small
        Three parallel token streams (strip, ADC, TDC) are embedded and fused with layer-relative positional and detector-layer embeddings. The incident photon momentum is projected by an MLP and prepended as a prefix token for global conditioning. The sequence is processed by 8 pre-norm transformer decoder blocks with 8-head causal self-attention and feed-forward networks ($512 \to 2048 \to 512$). Three output heads predict logits over the strip (412), TDC (2,005), and ADC (5,004) vocabularies. ADC uses soft section routing, blending PCAL, ECIN, and ECOUT heads via marginalized strip probabilities. At inference, tokens are sampled autoregressively; LAYER\_SEP marks layer transitions and EOS terminates the shower.}
    \label{fig:architecture}
\end{figure}

This section describes the architecture used in this work. 
We employ a decoder-only GPT-style autoregressive transformer (see, \textit{e.g.}, \cite{black2022gptneox20bopensourceautoregressivelanguage}) conditioned on the incident particle's momentum vector $\mathbf{p} = (p_x, p_y, p_z)$.
The overall workflow is summarized in Fig.~\ref{fig:architecture}, with additional details provided below.

\subsection*{Event Representation}

Each shower in the CLAS12 calorimeter consists of hits distributed across nine detector layers — three each in PCAL, ECIN, and ECOUT — with U, V, and W readout planes per calorimeter. Each hit carries three measurements: a strip index identifying the activated scintillator strip, a pulse-height in ADC counts, and a timing measurement in TDC counts.

Rather than representing an event as a fixed-dimensional vector or image, we encode it as a variable-length sequence of discrete tokens, enabling the use of an autoregressive language model to learn the conditional distribution over detector occupancy.

\subsection*{Tokenization}

We define three parallel integer token streams — strip, ADC, and TDC — that are aligned position-by-position and concatenated into a single embedding space at the input to the transformer. A special LAYER\_SEP token is inserted between each pair of adjacent detector layers (but not after the final layer before EOS), providing the model with an explicit structural signal for layer boundaries. Within each layer, hits are sorted by ascending strip index.
Each of the three streams carries its own SOS, EOS, LAYER\_SEP, and PAD tokens, but the streams are always aligned: position $i$ in the strip stream corresponds to the same hit as position $i$ in the ADC and TDC streams. The maximum sequence length is 190 positions (1 SOS + 9 layers $\times$ 20 hits + 8 SEP tokens + 1 EOS).

\subsubsection*{Strip Tokens}

Strip tokens encode the detector layer and strip identity jointly.
The strip tokens are assigned in contiguous, non-overlapping blocks of the vocabulary, one block per layer.
For layer $\ell$, let $N_\ell$ denote the number of strips in that layer, with
$N_\ell \in \{68,62,62,36,36,36,36,36,36\}$.
We define the layer offset as
$\mathrm{offset}_\ell = \sum_{k<\ell} N_k$,
so that strips in layer $\ell$ are assigned token indices
$
[\mathrm{offset}_\ell,\mathrm{offset}_\ell+N_\ell).
$
Thus, the detector layer is implicit in the token identity: given any strip token, the corresponding layer can be recovered by a precomputed lookup table. The strip vocabulary contains 412 tokens in total.

\subsubsection*{ADC Tokens}

Pulse-height values span roughly five orders of magnitude (1 to 94,920 ADC counts). To preserve resolution near the detection threshold without requiring an impractically large vocabulary, we employ a hybrid linear-log binning scheme with 5,000 bins. The lowest 50 bins are linearly spaced over $[1, 50]$, and the remaining 4,950 bins are logarithmically spaced over $[50, 94920]$. During decoding, bin centers are recovered using arithmetic midpoints for the linear region and geometric means for the log region. The ADC vocabulary contains 5,004 tokens (5,000 bins plus PAD, SOS, EOS, and LAYER\_SEP).

\subsubsection*{TDC Tokens}

Timing values (1 to 2,000,000 TDC counts) are logarithmically binned into 2,000 tokens. A dedicated TDC\_ZERO token is reserved for hits with no valid timing measurement. The TDC vocabulary contains 2,005 tokens.


\subsubsection*{Momentum Conditioning}

The momentum vector is projected into the model's embedding space by a
three-layer MLP with GELU activations whose final layer produces
$N_\text{layers}=9$ separate conditioning embeddings: 
\begin{equation}
\mathbf{E}_\text{mom} = \operatorname{reshape}\big(\text{MLP}(\mathbf{p})\big)
\in \mathbb{R}^{N_\text{layers} \times d_\text{model}} .
\end{equation}
These layer-specific momentum embeddings are prepended to the token sequence as a block of prefix tokens (one per detector layer) occupying positions $0$ through $8$, so that every subsequent hit position can attend to the full momentum context through the causal self-attention. Using one prefix token per layer, rather than a single shared token, provides a dedicated per-layer anchor for momentum conditioning.

\subsubsection*{Token Embedding Fusion}

Each of the three token streams is mapped to a learned embedding of dimension $d_\text{model}/3$, and the three are concatenated and linearly projected to the model dimension:
\begin{equation}
\mathbf{h} = \mathbf{W}_\text{proj}
\left[\, \mathbf{e}_\text{strip};~ \mathbf{e}_\text{ADC};~ \mathbf{e}_\text{TDC} \,\right],
\qquad
\mathbf{e}_\bullet \in \mathbb{R}^{d_\text{model}/3},\;
\mathbf{W}_\text{proj} \in \mathbb{R}^{d_\text{model} \times d_\text{model}} .
\end{equation}
Positional and layer-specific embeddings are then added,
\begin{equation}
\mathbf{h} = \mathbf{h} + \mathbf{e}_\text{pos} + \mathbf{e}_\text{layer},
\end{equation}
followed by layer normalization and dropout. Here $\mathbf{e}_\text{strip}$, $\mathbf{e}_\text{ADC}$, and $\mathbf{e}_\text{TDC}$ are learned lookup embeddings for each token stream, and $\mathbf{W}_\text{proj}$ is a learned projection that fuses them into the shared model space. The term $\mathbf{e}_\text{pos}$ is a learned position embedding indexed by the hit's rank within its current detector layer, reset to zero at each LAYER\_SEP, and $\mathbf{e}_\text{layer}$ is a learned embedding over the nine detector-layer indices plus one additional index for special tokens.

\subsubsection*{Transformer Decoder}

The model consists of 8 pre-norm transformer decoder blocks. Each block contains three main components: (1) an 8-head causal self-attention layer with head dimension 64, (2) a position-wise feed-forward network consisting of Linear($512 \to 2048$), GELU activation, Dropout, Linear($2048 \to 512$), and Dropout, and (3) pre-LayerNorm applied to both sub-layers with residual connections throughout.

\subsubsection*{Output Heads and Soft Section Routing}

Three separate linear projection heads predict the next token in each stream. The strip head projects from $\mathbb{R}^{512}$ to $\mathbb{R}^{412}$, the TDC head from $\mathbb{R}^{512}$ to $\mathbb{R}^{2005}$, and three section-specific ADC heads each project from $\mathbb{R}^{512}$ to $\mathbb{R}^{5004}$ (one for PCAL, one for ECIN, and one for ECOUT).

Because pulse-height distributions differ substantially between calorimeters, ADC prediction employs soft section routing rather than hard assignment. The three section-specific ADC logits are blended using the predicted strip probability marginalized over each section:
\begin{equation}
\hat{\mathbf{a}} = \pi_{\text{\tiny{PCAL}}} \mathbf{a}_\text{\tiny{PCAL}} + \pi_\text{\tiny{ECIN}} \mathbf{a}_\text{\tiny{ECIN}} + \pi_\text{\tiny{ECOUT}} \mathbf{a}_\text{\tiny{ECOUT}}
\end{equation}

where $\pi_s = \sum_{k \in \text{section } s} P(\text{strip}_k)$ is the predicted probability of a hit belonging to section $s$ and $\mathbf{a}$ represents the ADC output head. This differentiable routing eliminates any train/inference mismatch that would arise from hard argmax-based routing.

\subsubsection*{Training Objective}

The model is trained by minimizing a composite loss combining the per-stream
token likelihoods with several auxiliary terms that regularize sequence structure
and occupancy:
\begin{equation}
\mathcal{L} = \mathcal{L}_\text{strip} + \mathcal{L}_\text{ADC}
            + \mathcal{L}_\text{TDC}
            + \lambda_\text{occ}\,\mathcal{L}_\text{occ}
            + \lambda_\text{len}\,\mathcal{L}_\text{len}
            + \lambda_\text{prox}\,\mathcal{L}_\text{prox} .
\end{equation}
The primary training objective is the token-level cross-entropy of the three
output streams, $\mathcal{L}_\text{strip}+\mathcal{L}_\text{ADC}+\mathcal{L}_\text{TDC}$,
\textit{i.e.},\ the standard autoregressive likelihood. Because the structural tokens that
delimit the sequence (LAYER\_SEP, EOS) are rare relative to hit tokens, they are
modestly up-weighted within the strip cross-entropy to correct this class
imbalance, so that layer boundaries and sequence termination are learned as
reliably as the hits themselves.
The bare token likelihood does not, however, directly constrain the global structure of an event---how many hits each layer should contain and when the shower should end. We therefore add three lightweight auxiliary regularizers, each targeting a specific structural property. An occupancy term, $\mathcal{L}_\text{occ}$, discourages the model from assigning probability to termination tokens at positions that should contain a genuine hit (mitigating premature termination); a per-layer length term, $\mathcal{L}_\text{len}$, matches the expected number of emitted hits to the true per-layer count via a robust Huber loss; and a differentiable proxy term, $\mathcal{L}_\text{prox}$, nudges the generated marginal distributions (sequence length, and the per-layer strip and ADC occupancy histograms) toward the data. These terms address exactly the multiplicity- and occupancy-level behavior evaluated in Section~\ref{sec:results}. All loss weights are fixed across every experiment reported here and are listed in Table~\ref{tab:hparams}; the model is not sensitive to their precise values.

We found these structural regularizers to be necessary rather than merely beneficial. Trained on the token cross-entropy alone, the model exploits the likelihood by terminating sequences prematurely: it emits the EOS and LAYER\_SEP tokens far too early, collapsing toward short, under-occupied events with a mean of only $\sim$10 hits per event (versus 36 in data) and severely depressed occupancy in the outer calorimeter layers. The occupancy and length terms penalize exactly this behavior, and restore agreement with the true multiplicity and occupancy distributions reported in Section~\ref{sec:results}.

\begin{table}[h!]
  \centering
  \caption{Loss-term weights, held fixed across all experiments.}
  \label{tab:hparams}
  \begin{tabular}{lc}
    \toprule
    Term & Weight \\
    \midrule
    Strip / ADC / TDC cross-entropy        & $1.0$ \\
    \quad LAYER\_SEP up-weight (strip)      & $4.5$ \\
    \quad EOS up-weight (strip)             & $3.0$ \\
    Occupancy regularizer                   & $3.0$ \\
    Per-layer length (Huber, $\delta=3$)    & $0.4$ \\
    Distribution-matching proxy             & $0.1$ \\
    \bottomrule
  \end{tabular}
\end{table}

\subsubsection*{Optimization}

Training uses AdamW with a learning rate of $4 \times 10^{-4}$ under a cosine annealing schedule featuring 1,000 warmup steps and a minimum LR ratio of 0.1. Weight decay is set to $10^{-2}$, gradients are clipped at 1.0, and the batch size is 400. The model is trained for up to 100,000 steps.

\section{Results}\label{sec:results}

\begin{figure}[h!]
    \centering
    \includegraphics[width=0.7\textwidth]{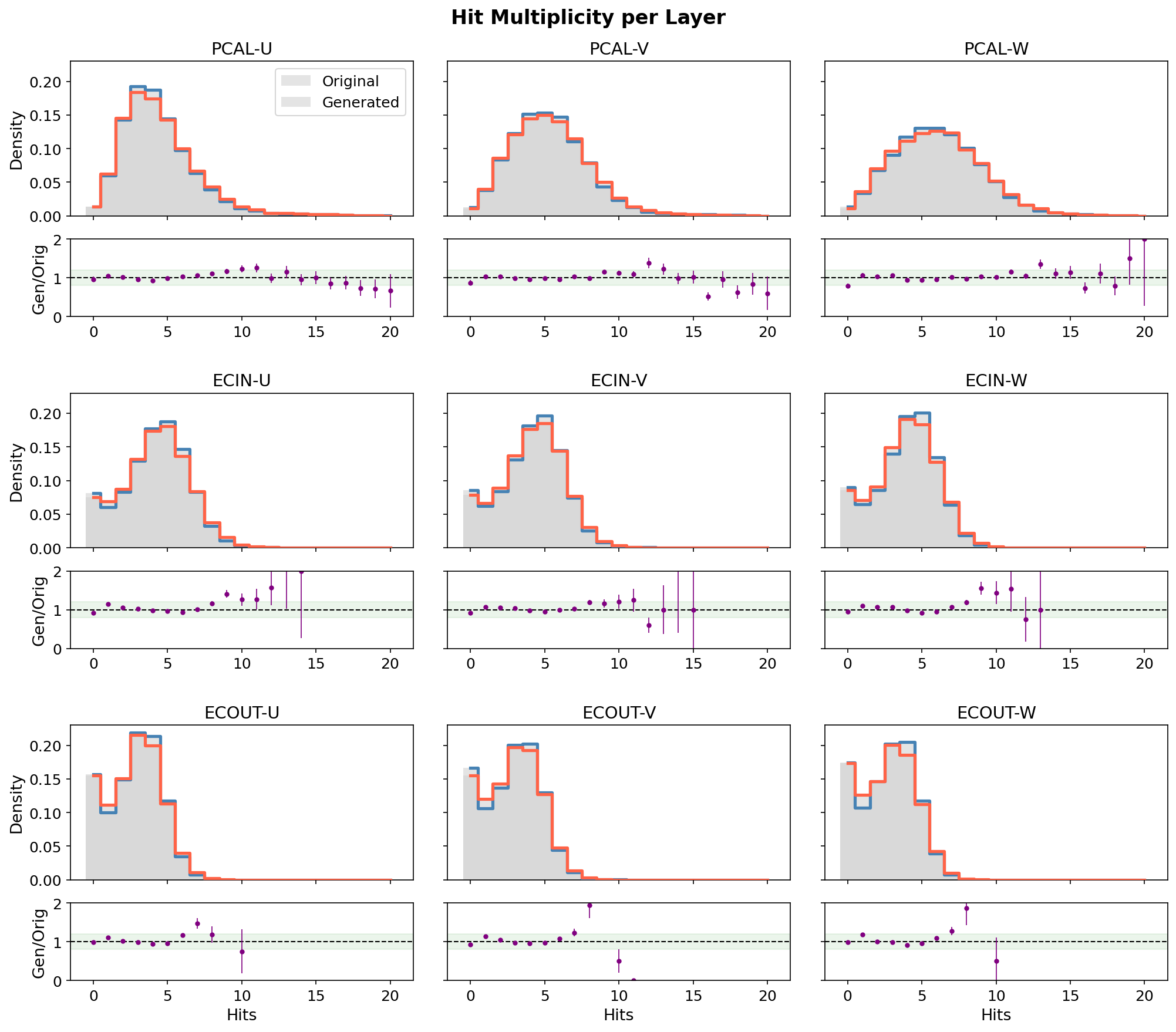}
    \caption{Per-layer hit multiplicity (number of fired strips per event) for
        each of the nine CLAS12 calorimeter layers (PCAL, ECIN, ECOUT in the U/V/W
        views). Grey histograms and the overlaid step curves compare the original
        data (blue) with events sampled from the autoregressive generator (red),
        both normalised to unit area. The lower panel of each cell shows the
        generated-to-original ratio with statistical error bars; the dashed line
        marks unity and the green band the $\pm20\%$ region.}    
    \label{fig:multiplicity}
\end{figure}

In this section, we evaluate the model by generating events conditioned on original photon momenta drawn from the test set and comparing the resulting distributions to \textsc{Geant4} ground truth. 
For each observable, we overlay the generated and original distributions and show their bin-by-bin ratio, together with a fixed $\pm 20\%$ reference band around unity (see Figures \ref{fig:multiplicity} to \ref{fig:total_adc}). This band provides a visual guide for assessing the level of agreement.

To summarize agreement with a single interpretable number per observable, we use different metrics suited to each quantity's scale and units. For shower position and hit multiplicity, we report the Wasserstein-1 (earth-mover's) distance computed on raw samples, which carries the  physical units of the observable (strips and hits, respectively). For  per-hit ADC, where values span several orders of magnitude, we compute  the Wasserstein-1 distance in $\log_{10}(\text{ADC})$ space, averaged across all nine calorimeter layers. For the energy scale, we report the  ratio of mean total deposited ADC between generated and original events,  averaged across momentum bins, which directly quantifies any systematic  offset in the overall energy response. These metrics are summarized in  Table~\ref{tab:wass}.
More details are provided in the following.

\subsection*{Hit Multiplicity and Sequence Length}

\begin{figure}[h!]
    \centering
    \includegraphics[width=0.8\textwidth]{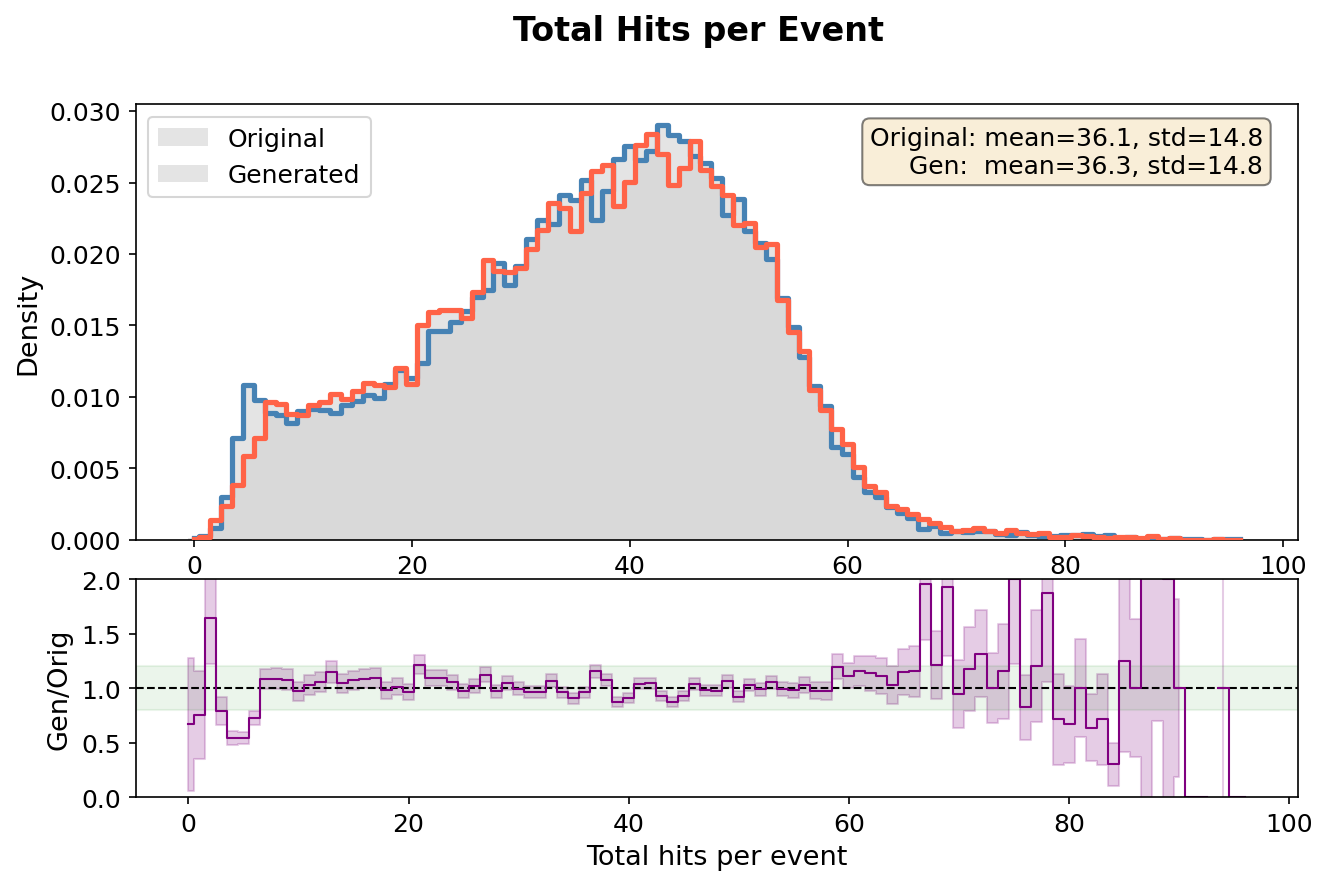}
    \caption{Total number of hits per event (summed over all nine layers) for
    original (blue) and generated (red) events, normalised to unit area, with
    the corresponding means and standard deviations annotated. The lower panel
    shows the generated-to-original ratio (unity dashed; $\pm20\%$ band in
    green).}
    \label{fig:seqlength}
\end{figure}

Figure~\ref{fig:multiplicity} shows the per-layer hit multiplicity
distributions. The model reproduces the peak position, the falling tail, and the
per-layer mean across all nine layers, with generated-to-original ratios lying
within the $\pm20\%$ band over the populated range. The global hit-count
distribution (Figure~\ref{fig:seqlength}) is reproduced equally well: the
generated mean of $36.3\pm14.8$ hits per event matches the original mean of
$36.1\pm14.8$ to better than one hit, with the generator marginally over-filling
the high-multiplicity tail.

\subsection*{Layer Occupancy}
\FloatBarrier

\begin{figure}[h!]
    \centering
    \includegraphics[width=0.8\textwidth]{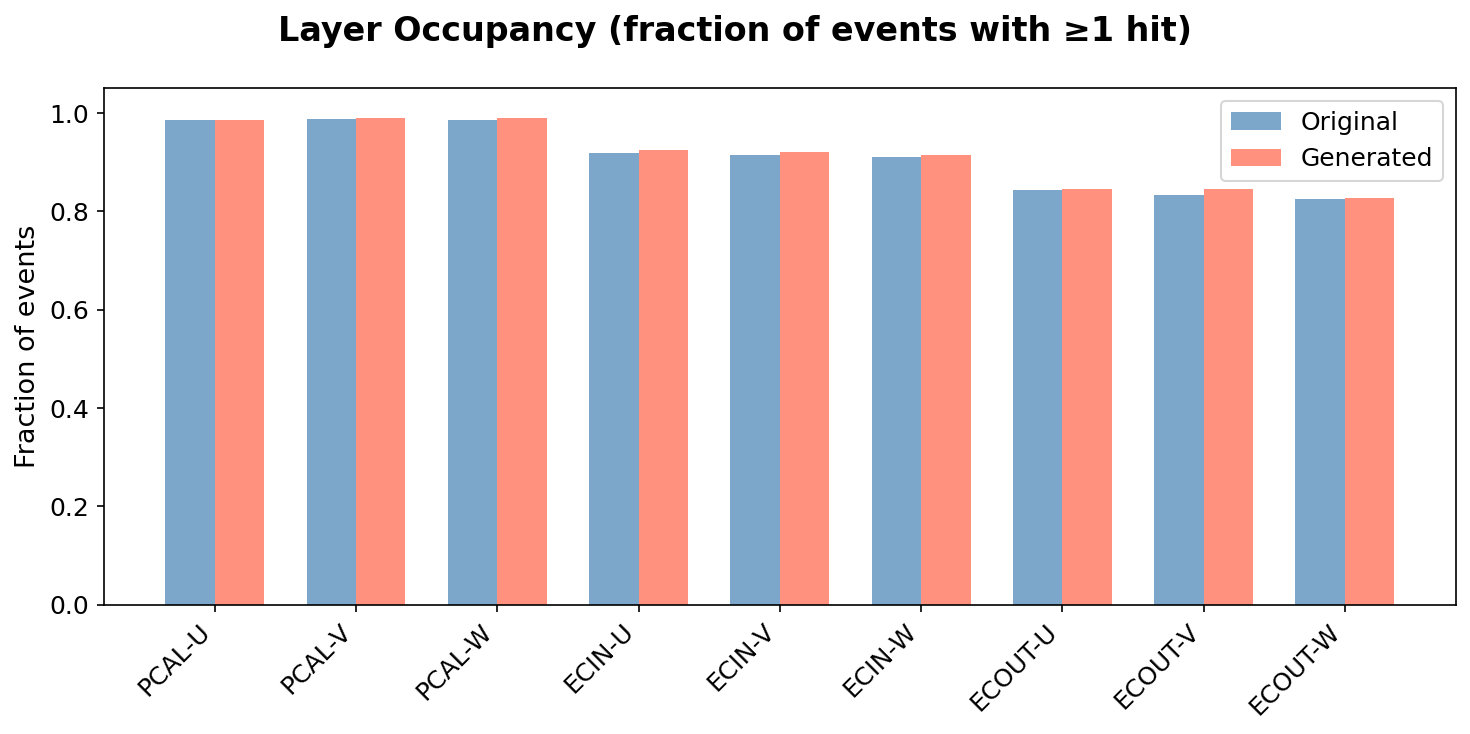}
    \caption{Layer occupancy, defined as the fraction of events with at least one
    hit in each of the nine calorimeter layers, comparing original (blue) and
    generated (red) events. This summarises how reliably the generator activates
    each layer.}
    \label{fig:occupancy}
\end{figure}

Figure~\ref{fig:occupancy} shows the fraction of events with at least one
reconstructed hit per layer. Occupancy is reproduced across all nine layers to
within roughly one percentage point: the PCAL layers are activated in
$\sim$98--99\% of events, the ECIN layers in $\sim$91--92\%, and the ECOUT layers
in $\sim$83--85\%, in both data and generation. The model thus correctly learns
the decreasing activation probability through the depth of the calorimeter
without systematic over- or under-firing.

\FloatBarrier
\subsection*{Strip Distributions}

\begin{figure}[!]
    \centering
    \includegraphics[width=0.725\textwidth]{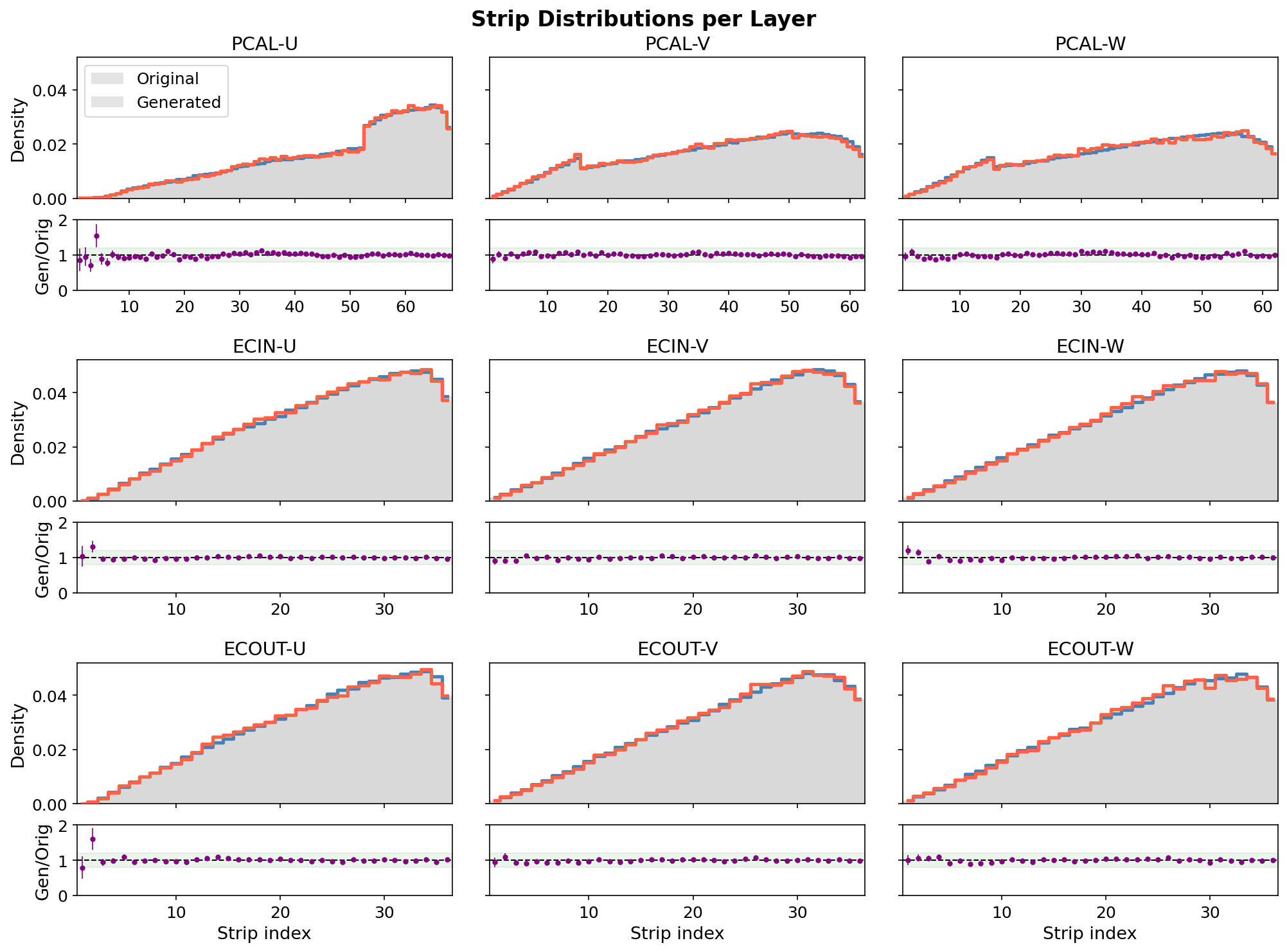}
   \caption{
        Fired strip indices per layer for original (blue) and generated (red) events, normalized to unit area, testing whether the generator reproduces each calorimeter view's spatial occupancy. Lower panels show generated-to-original density ratios with statistical uncertainties; dashed lines mark unity, green bands $\pm20\%$.
        }
    \label{fig:strips}
\end{figure}

\begin{figure}[!]
    \centering
    \includegraphics[width=0.725\textwidth]{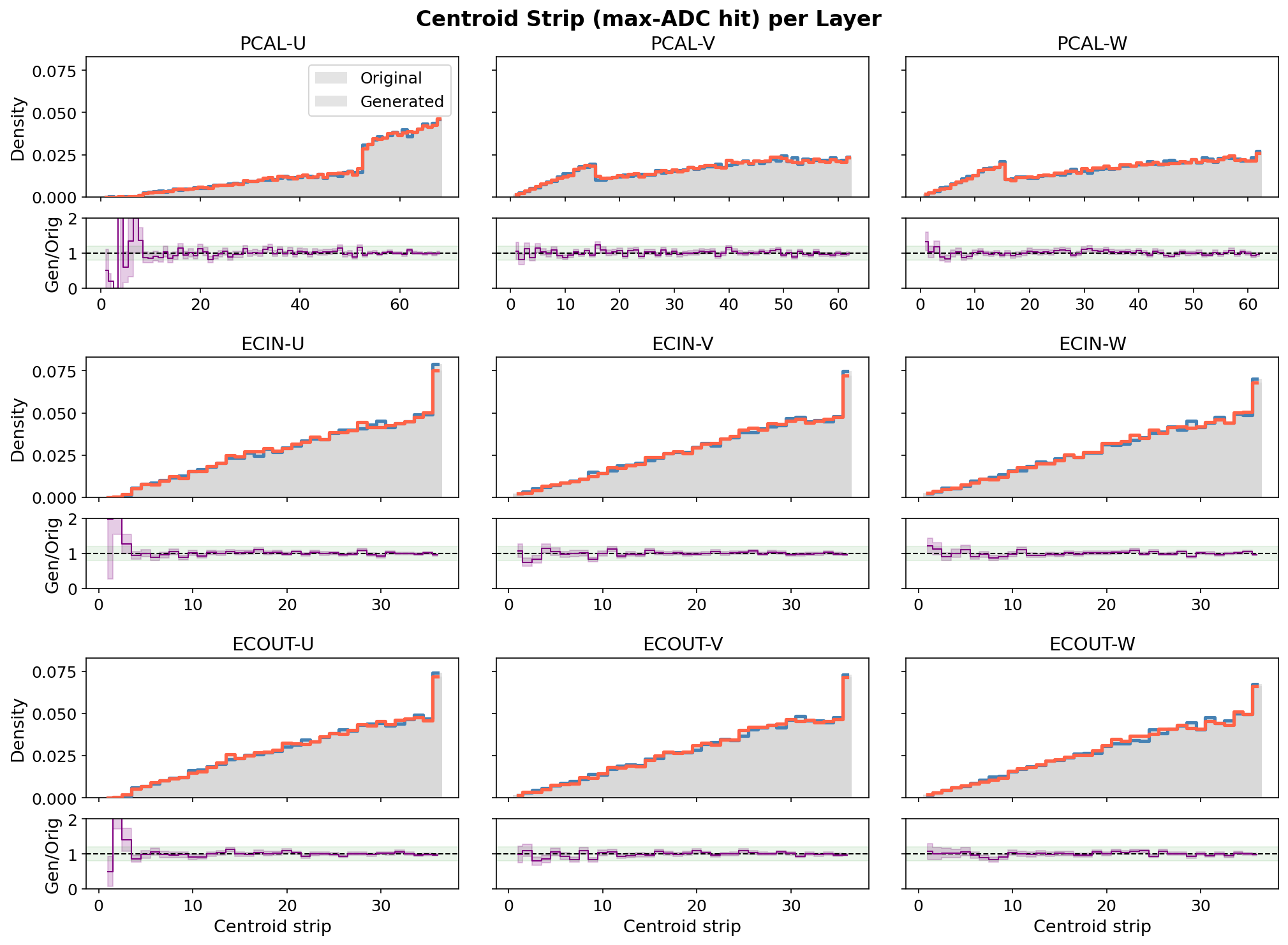}
    \caption{
        Centroid strip distributions per layer, defined by the highest-ADC hit, for original (blue) and generated (red) events, normalized to unit area, testing whether the generator places the shower core correctly within each view. Lower panels show generated-to-original density ratios; dashed lines mark unity, green bands $\pm20\%$.
    }
    \label{fig:centroid}
\end{figure}

The marginal strip-index distributions are the most accurately reproduced (Figure~\ref{fig:strips}): the characteristic rising profile toward
higher strip indices, reflecting the geometry of the CLAS12 forward
calorimeters, is correctly learned in every layer, with generated-to-original
ratios near unity across the full strip range. The mean shower position agrees
with the data to within $0.12$ strips, comparable to, and in fact slightly
below, the $0.15$-strip noise floor obtained from two independent original
samples, \textit{i.e.}, statistically indistinguishable agreement. The centroid strip
distributions (the strip carrying the maximum ADC value per layer) are similarly
well reproduced (Figure~\ref{fig:centroid}), confirming that the model places the
shower core at the correct location within each view.

\subsection*{ADC Distributions}

\begin{figure}[!]
    \centering
    \includegraphics[width=0.725\textwidth]{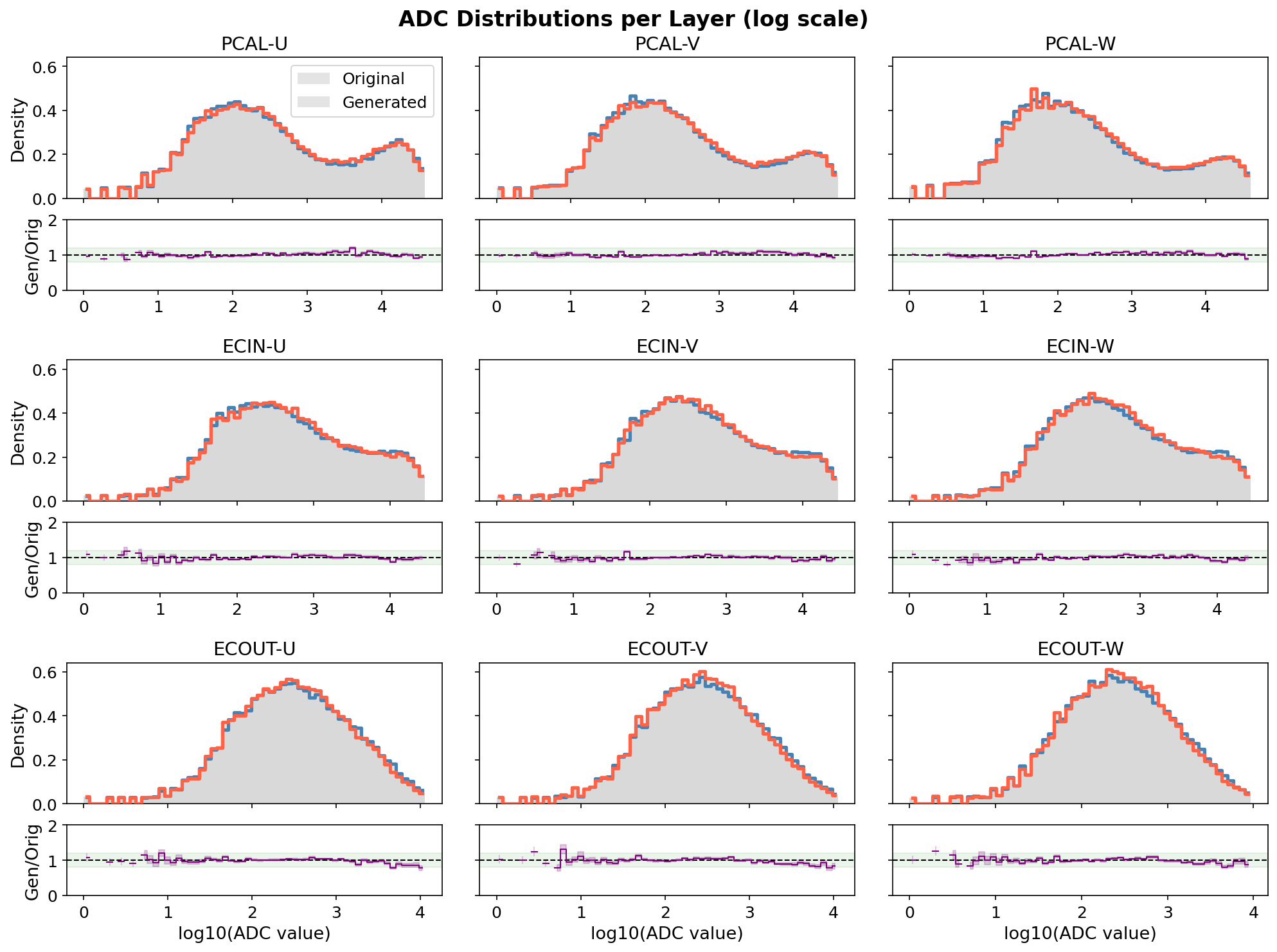}
    \caption{
        Per-hit ADC distributions per layer, shown as $\log_{10}(\mathrm{ADC})$ for non-zero hits, for original (blue) and generated (red) events, normalized to unit area. Lower panels show generated-to-original density ratios with statistical uncertainties; dashed lines mark unity, green bands $\pm20\%$.}
    \label{fig:adc}
\end{figure}

\begin{figure}[!]
    \centering
    \includegraphics[width=0.725\textwidth]{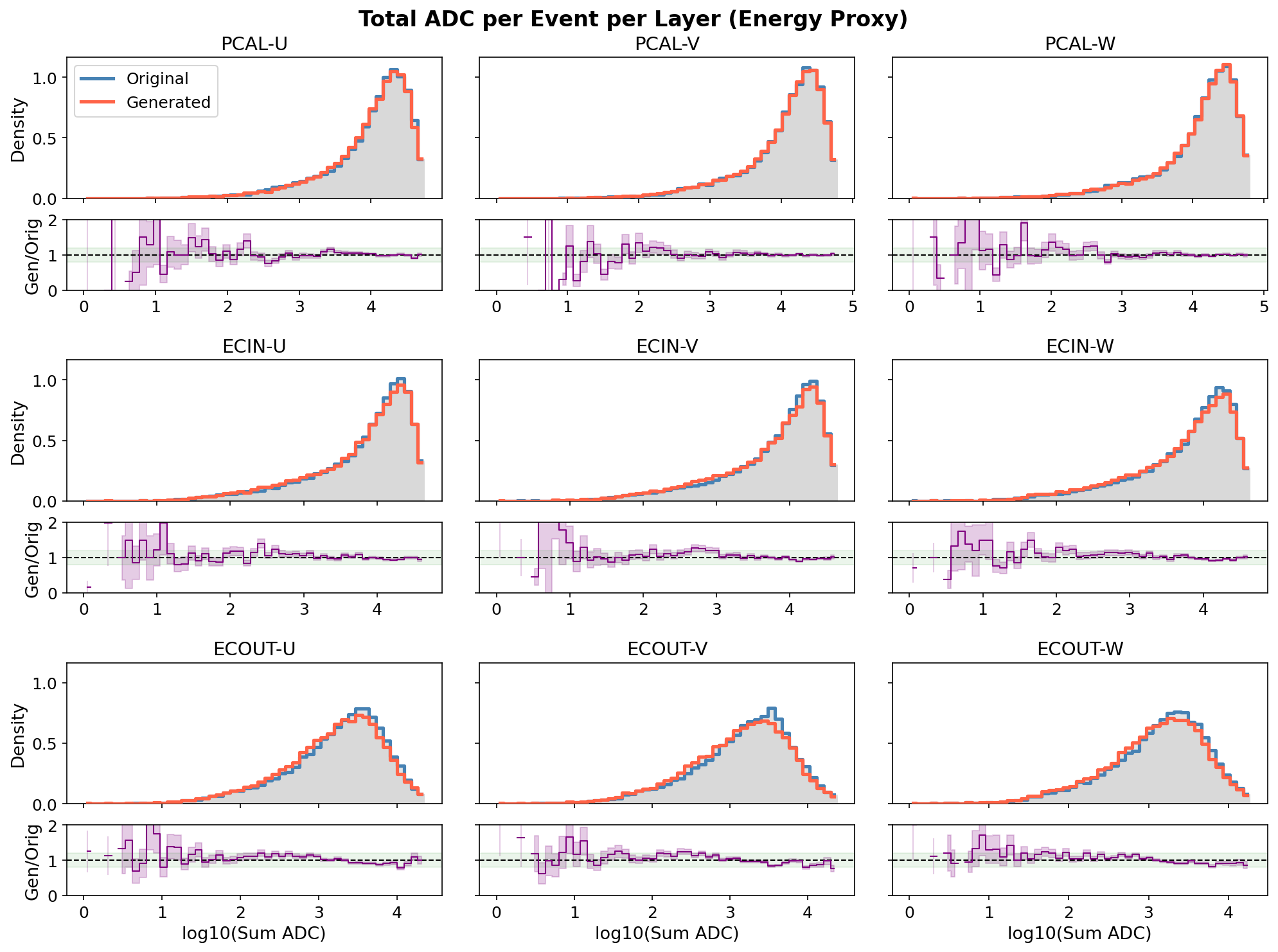}
    \caption{
        Total ADC per event in each layer, used as an energy proxy and shown as $\log_{10}(\sum \mathrm{ADC})$ for non-zero deposits, for original (blue) and generated (red) events, normalized to unit area. Lower panels show generated-to-original density ratios with statistical uncertainties; dashed lines mark unity, green bands $\pm20\%$.
        }
    \label{fig:total_adc}
\end{figure}

The pulse-height distributions span several orders of magnitude in ADC counts
and are reproduced with high fidelity (Figure~\ref{fig:adc}). The model captures
the characteristic bimodal structure in the PCAL layers and the broader
single-peak distributions in the ECIN and ECOUT layers, with
generated-to-original ratios near unity throughout. The total ADC per event per
layer (Figure~\ref{fig:total_adc}) is likewise well described. In aggregate, the
per-hit and total-ADC distributions agree with the data to within $3\%$ with the generator
sitting marginally high in the total-ADC tail.

\FloatBarrier
\subsection*{Calorimeter Energy--Momentum Response}
\label{sec:calo_energy}

\begin{figure}[htbp]
  \centering
  \includegraphics[width=0.8\textwidth]{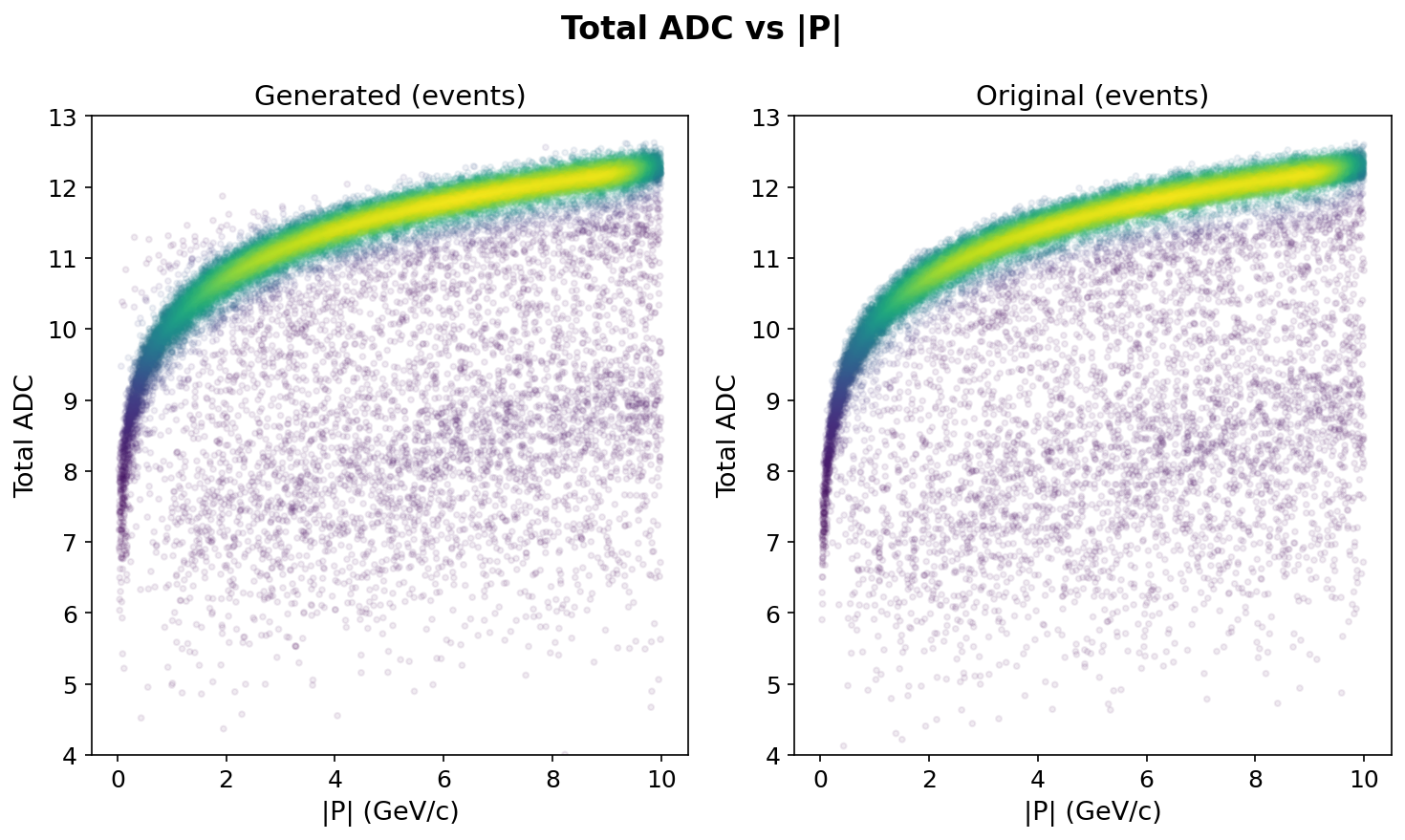}
  \caption{Total ADC per event (on a natural-log scale) versus incident momentum
    magnitude $|P|$, with point color encoding the local density (Gaussian
    kernel-density estimate). The left panel shows generated events and the right
    panel the original data, providing a direct comparison of the calorimeter
    energy--momentum response reproduced by the generator.}
  \label{fig:total_adc_vs_momentum}
\end{figure}

A key test of the momentum conditioning is whether the generator reproduces the
correlation between incident particle momentum and total deposited energy, rather
than merely matching the marginal energy distribution of
Figure~\ref{fig:total_adc}. Figure~\ref{fig:total_adc_vs_momentum} compares the
joint distribution of total ADC per event against $|P|$ for generated (left) and
original (right) events. 
The mean total ADC matches the data to within 1--2\% across all $|P|$ bins, and the event-to-event energy spread is reproduced to within a few percent, demonstrating that the model has learned the full energy--momentum response of the calorimeter rather than merely the marginal energy distribution.

\subsection*{Aggregate Performance}
\FloatBarrier

Across every observable the generated distributions track the \textsc{Geant4}
ground truth closely, with generated-to-original ratios remaining within the
$\pm20\%$ band over the populated range of each figure. Table~\ref{tab:wass}
summarizes the level of agreement in the physical units of each observable,
alongside the statistical noise floor set by two independent original samples.
The shower position is reproduced at the level of the noise floor and the hit
count to within a fraction of a hit per event. The calorimeter energy response is
reproduced with comparable fidelity: the total deposited energy is correct to
within 1--2\% in absolute scale, the per-hit pulse height to a few percent, and
the energy resolution and its dependence on incident momentum are likewise
reproduced to within a few percent as shown in Fig. \ref{fig:total_adc_vs_momentum}.
Together with the correctly reproduced layer occupancy, these results demonstrate
that the model reproduces the calorimeter response with high fidelity. The small
residual over-production of hits and energy is the focus of ongoing optimization
of the loss weighting and sampling temperature.

\begin{table}[h!]
  \centering
  \caption{Agreement between generated and original (\textsc{Geant4})
    distributions, summarised by a single metric per observable. Shower position and hits per event are reported as Wasserstein-1 (earth-mover's) distances in physical units, computed directly on raw samples. For per-hit ADC, where values span several orders of magnitude, we compute the Wasserstein-1 distance in $\log_{10}(\text{ADC})$ space, averaged across all nine calorimeter layers; a value of $0.017$ $\log_{10}(\text{ADC counts})$ indicates close agreement between the generated and original per-hit energy deposit distributions. The energy-scale entry is the ratio of mean total deposited ADC per event between generated and original distributions, averaged across momentum bins; a value of $1$--$2\%$ indicates the mean energy response is reproduced to within $1$--$2\%$ across the full momentum range.}
  \label{tab:wass}
  \begin{tabular}{lcc}
    \hline
    Observable                       & Gen vs.\ \textsc{Geant4} \\
    \hline
    Shower position                  & $0.12$ strips\\
    Hits per event                   & $0.27$ hits\\
    Per-hit ADC                      & $0.017 \log(ADC)$ \\
    Energy scale (mean total ADC)    & $1$--$2\%$ \\
    \hline
  \end{tabular}
\end{table}

\FloatBarrier

\section{Inference}\label{sec:inference}

\subsection{Inference Procedure}
 
At inference time, the model operates autoregressively to generate detector hits for a given incident photon momentum. The procedure begins with the momentum vector $\mathbf{p} = (p_x, p_y, p_z)$, which is embedded via the MLP and prepended as a context token. The model then generates tokens sequentially across the three parallel streams (strip, ADC, TDC) until termination, with layer boundaries marked by LAYER\_SEP tokens and the end-of-sequence (EOS) token signaling the completion of the shower.
 
\subsection{Token Sampling Strategy}
 
During generation, tokens are sampled from the output logits using temperature-controlled top-$k$ sampling to balance diversity and quality. In the results presented here, temperature is set to $1.0$ and top-$k$ is set to $20$ (drawing samples from the top 20 most probable tokens at each step). These hyperparameters were chosen to mitigate mode collapse while maintaining physical realism; ongoing tuning of these parameters is a focus of future optimization work.
 
The causal self-attention mask ensures that generation at each position depends only on previously generated tokens and the momentum conditioning, preserving the autoregressive structure. No teacher forcing or external guidance is applied during inference---all generations are driven purely by the learned conditional probability distribution.
 
\subsection{Handling Variable-Length Outputs}
 
Because electromagnetic showers exhibit variable multiplicity across detector layers and across events, the model naturally produces variable-length sequences. The maximum sequence length of 190 tokens accommodates the theoretical maximum of 20 hits per layer across 9 layers, plus structural tokens (SOS, LAYER\_SEP, EOS). In practice, most events terminate well before this maximum, with a mean of $36.3 \pm 14.8$ hits per event as shown in Figure~\ref{fig:seqlength}. Early termination via the EOS token provides a clean stop criterion without requiring external post-processing or truncation.

\subsection{Key--Value Caching}

To make autoregressive generation efficient we adopt key--value (KV) caching, the
standard inference-time optimization for decoder-only transformers in large
language models~\cite{kvcache}. Without it, the self-attention keys and values for
the entire preceding sequence are recomputed at every step, even though they are
unchanged; caching them reduces the cost per generated token from
$\mathcal{O}(T^2)$ to $\mathcal{O}(T)$ in the sequence length $T$.

Generation proceeds in two phases. In a prefill pass the momentum context
token and the start-of-sequence token are processed together, populating the
key--value cache of every decoder block. Each subsequent step then embeds only
the single newest hit---formed jointly from its strip, ADC and TDC tokens---and
appends its key and value to the cache, so that the new token attends over the
full history without recomputation. Because the three token streams are fused
into one hidden state per sequence position, a single cache per block serves all
of them. This optimization, together with the batched sampling described above,
underlies the throughput reported in Table~\ref{tab:inference_l40s}.

\subsection{Computational Performance}
 
Inference throughput was measured on a single NVIDIA L40S GPU (48 GB VRAM) using the model configuration described in Section~\ref{sec:architecture}. Table~\ref{tab:inference_l40s} reports the generation rate across a range of batch sizes, measured using production-like sampling parameters (temperature $= 1.0$, top-$k$ $= 20$, maximum 190 steps).
 
The model achieves throughput exceeding $700$ events per second at batch size 256, with peak performance of $763.2$ events/s at batch size 1024. This corresponds to an average latency of $1.31$ ms per event. Throughput scales efficiently with batch size up to 1024, with modest degradation at batch size 4096 due to memory pressure on the GPU.
For comparison, traditional \textsc{Geant4}-based~\cite{geant4} CLAS12 simulation typically requires $10$--$100$ seconds of CPU time per event depending on complexity and system load, representing a speedup of approximately $10^4$--$10^5\times$ over conventional full simulation. Even accounting for the modest underfilling of hit multiplicity discussed in Section~\ref{sec:results}, this represents a transformative reduction in computational cost for large-scale experimental programs.
 
The inference code is deterministic and fully reproducible given a fixed random seed. Deployment in a production setting would require minimal additional infrastructure beyond standard PyTorch serving tools, making rapid integration with existing reconstruction pipelines feasible.

\begin{table}[t]
  \centering
  \caption{Autoregressive inference throughput of our GPT-style model
           on an NVIDIA L40S (38.4\,M parameters; $d_\text{model}=512$,
           8 layers, 8 heads; temperature $=1.0$, top-$k=20$, max.\ 190 steps;
           20{,}480 events per batch size).}
  \label{tab:inference_l40s}
  \begin{tabular}{rrrr}
    \toprule
    Batch & Events/s & ms/event & Tokens/s \\
    \midrule
     256 & 700.3 & 1.428 & 62\,906 \\
     512 & 748.5 & 1.336 & 69\,592 \\
    1024 & 763.2 & 1.310 & 73\,952 \\
    2048 & \textbf{778.7} & \textbf{1.284} & \textbf{77\,407} \\
    4096 & 724.7 & 1.380 & 74\,792 \\
    \bottomrule
  \end{tabular}
\end{table}

\section{Summary and Conclusion}
We have presented a GPT-style autoregressive transformer as a fast surrogate model for the CLAS12 electromagnetic calorimeter, trained to generate realistic detector hits conditioned on incident photon momentum. By encoding calorimeter events as variable-length sequences of discrete strip, ADC, and TDC tokens across all nine readout layers, the model naturally captures the correlations between hit multiplicity, spatial position, and energy deposition that characterize real photon-induced electromagnetic showers.

The generator reproduces the \textsc{Geant4} ground truth with high fidelity across all evaluated observables. Hit multiplicity, strip distributions, and layer occupancy are recovered to within the statistical noise floor in several cases, while ADC distributions agree to within a few percent. The detector energy-response dependence on incident momentum is also reproduced within $1$–$2\%$ across all $|P|$ bins.
Inference throughput exceeds 700 events per second on a single NVIDIA L40S GPU, representing a speedup of approximately $10^4$--$10^5$ over traditional \textsc{Geant4}-based simulation while maintaining the physics fidelity required for high-luminosity experimental programs.

Several directions remain open for future work. Ongoing optimization of the loss weighting and sampling temperature aims to reduce the residual over-production of hits and energy in the high-multiplicity tail. Extension of the model to additional particle species and to other CLAS12 detector subsystems would broaden its applicability within the experimental program. Longer term, the autoregressive tokenization framework introduced here is general enough to accommodate other strip-based or segmented detectors, suggesting a path toward a unified generative model for multi-subsystem simulation.

Together, these results demonstrate that autoregressive language model architectures, adapted to the discrete, sequential structure of detector readout data, offer a compelling and practical alternative to traditional Monte Carlo methods for fast calorimeter simulation.


\section*{Data availability statement}
The data that support the findings of this study are available upon reasonable request from the authors.

\section*{Code Availability}
The code is publicly available at \href{https://github.com/wmdataphys/CLAS12_fast_sim}{https://github.com/wmdataphys/CLAS12\_fast\_sim}.


\section*{Acknowledgments}
%
%
This material is based upon work supported by the National Science Foundation under Grant No. 2443510. 
The William \& Mary group acknowledges support from this grant.
The authors acknowledge William \& Mary  Research Computing for providing computational resources and technical support that have contributed to the results reported within this article.
This material is also based upon work supported by the U.K. Science and Technology Facilities Council under grants ST/Z510312/1 and ST/Y000315/1.


\section*{References}
\bibliographystyle{iopart-num}
\bibliography{biblio}

\clearpage

\end{document}